\documentstyle[11pt,paspconf,psfig]{article}

\def\halve{{\scriptstyle 1 \over 2}}

\begin{document}

\title{Galaxy formation in dissipationless N-body models}
\author{Eelco van Kampen}
\affil{Theoretical Astrophysics Center, Juliane Maries Vej 30,
DK-2100 Copenhagen, Denmark, and Royal Observatory Edinburgh, Blackford Hill,
Edinburgh EH9 3HJ, U.K.}

\begin{abstract}
We present a method of including galaxy formation in dissipationless N-body
simulations. Galaxies that form during the evolution are identified
at several epochs and replaced by single massive soft particles. This allows
one to produce two-component models containing galaxies and a background
dark matter distribution. We applied this technique to obtain two sets of 
models: one for field galaxies and one for galaxy clusters. 
We tested the method for the standard CDM scenario for structure formation
in the universe. A direct comparison of the simulated galaxy distribution
to the observed one sets the amplitude of the initial density fluctuation
spectrum, and thus the present time in the simulations.
The rates of formation and merging compare very well to simulations that
include hydrodynamics, and are compatible with observations.
We also discuss the cluster luminosity function.
\end{abstract}


\section{Introduction}

In order to test models for structure formation in the Universe it
is necessary to pinpoint galaxies within the modelled large-scale distribution
of matter. As this distribution evolves in a non-linear fashion, the use
of N-body methods is usually required.
Galaxies should form, under the influence of gravity, within an N-body
simulation in a self-consistent way. However, such simulations suffer from
a numerical problem: small groups of particles that represent galaxies
get disrupted by numerical two-body effects within clusters (Carlberg 1994;
van Kampen 1995). This can be solved by replacing each group of particles by a
single `galaxy particle' just after they have formed into a virialised system
that resembles a galactic halo, thus ensuring their survival.
This should also produce galaxies at the right time and the right
place, with a spectrum of masses. In van Kampen (1995) such galaxy particles
were only formed at a single epoch. Here we extend this scheme to `continuous'
galaxy formation by applying the algorithm several times during the evolution.
This means that merging of already-formed galaxies is taken into account
as well, although only in a schematic fashion.
An important advantage of having a galaxy formation algorithm added to the
N-body integrator is that the time normalisation is now intrinsically fixed,
since one can {\it directly} compare the properties of the distribution of
simulated galaxies to those of the observed galaxy distribution.

\section{Galaxy formation recipe}

\subsection{Outline}

For the identification of galaxies during the evolution of the large-scale
matter distribution we use the {\it local density percolation} algorithm
(see also van Kampen 1995). Instead of a fixed linking
length, we modulate the linking length according to the {\it local density}\
around each particle in such a way that the linking length is shorter in
high-density environments. This partly resolves the cloud-in-cloud problem.
We form (and merge) galaxies several times during the evolution,
so we have to consider percolation of {\it unequal}\ mass particles.
In the following ${\bf x}$ and ${\bf v}$ are {\it comoving} variables.

\subsection{Local density percolation for unequal mass particles}

In the local density percolation scheme particles are linked together if they
are separated by a certain fixed fraction $p$ of the Poissonian average nearest
neighbour distance $x^{\rm P}_{\rm nn}\equiv [4\pi\langle n\rangle /3]^{-1/3}$,
where $\langle n\rangle$ is the mean number density of particles, 
modulated by the local number density $n^{\rm G}({\bf x},s)$, which is
$n({\bf x})$ Gaussian smoothed at the scale $s\ x^{\rm P}_{\rm nn}$.
Thus, $p$ and $s$ are the (dimensionless) free parameters of the algorithm.
All models in this paper have $\langle n \rangle=8.0\ h_0^3$Mpc$^{-3}$,
so $x^{\rm P}_{\rm nn}=0.31 h_0^{-1}$Mpc. For unequal mass particles we need
an extra modulation according to their masses. Initially, all particles have
mass $m_0$. As galaxies form, particles arise with masses $m_i$ that are
integer multiples of $m_0$. Galaxy particles with mass $m_i$ that are put
$\sqrt{m_i/m_0}$ times further away will exert the same gravitational
force as dark particles with mass $m_0$. So we should take the factor
$\sqrt{m_i/m_0}$ as the second modulation factor.
This gives a local percolation length
$$ R_{\rm p}({\bf x}_i,p,s) = p\ x^{\rm P}_{\rm nn} \sqrt{m_i\over m_0}
   \Bigl[{n^{\rm G}({\bf x}_i,s)\over\langle n\rangle}\Bigr]^{-{1/3}}\ 
  .\eqno(1)$$
Since each particle has its own linking length, we use their mean to test
pairs of particles. Furthermore, to prevent excessive percolation lengths,
we adopt an absolute maximum of $x^{\rm P}_{\rm nn}/2$ for the pair linking
lengths (after taking the mean of the individual ones), i.e.\ a lower limit
for the local particle density which is equal to eight times the background
particle density.

In addition, we require pairs to have a relative pairwise velocity
$$ v_{||} \equiv({\bf v}_j - {\bf v}_i)\cdot({\bf x}_j - {\bf x}_i)
   / |{\bf x}_j - {\bf x}_i| \eqno(2) $$
of less then 800 km s$^{-1}$.
This is twice the relative pairwise velocity dispersion
at separations around $x^{\rm P}_{\rm nn}$ for the field, and somewhat smaller
than found for a sample including the Coma cluster (Mo, Jing \& B\"orner 1993).
It is also twice the maximum internal velocity dispersion we allow for a group.
We include this velocity linking length to exclude fast-moving particles which
are geometrically linked to a group. This often occurs within the potential
wells of galaxy clusters. One should see this criterion as the velocity
equivalent of a (constant) spatial linking length, so that we actually
find groups in phase-space.
However, the velocity linking length is less restrictive than
the spatial linking length since it serves a different purpose, as said.

\subsection{Virial equilibrium criterion for unequal mass particles}

A group of particles should only be transformed into a single, soft
galaxy particle if it forms a physical system roughly in virial equilibrium.
This `virial criterion' is a necessary addition to the local density
percolation algorithm for the purpose of defining galaxies. We will
use a virial equilibrium criterion in a simplified form using the half-mass
radius $R_{\rm h}$, motivated by Spitzer (1969) who found that for many
equilibrium systems the virial equilibrium equation can be written as
(where $v$ is now the proper velocity)
$$\sigma_v\equiv\langle v^2 \rangle \approx 0.4 {GM\over R_{\rm h}}\ .
  \eqno(3)$$
If galaxies are identified only once, one has to deal with equal mass
particles, and $R_{\rm h}$ can simply be calculated by obtaining the
median of the distances of all particles with respect to the centre of
the group being tested. For groups where the masses of the member particles
can differ a few orders of magnitude, the half-mass radius will often
exactly coincide with a galaxy particle.
This makes the median (i.e.\ the half-mass radius) a rather noisy estimater
for the total gravitational energy of a group of unequal mass particles.
Because for many probability distributions the mean and the median are almost
identical, we use the mass-weighted mean distance from the centre of the group
as an estimator for $R_{\rm h}$. This is a more smoothly-defined and
well-behaved quantity than the median distance. The new galaxy particle has
a softening parameter corresponding to the $R_{\rm m}$ of the original
group, which certifies a reasonable conservation of energy
(see van Kampen 1995).

Discreteness noise will cause some scatter in the group quantities, so we
should allow for some tolerance in the difference between the estimated virial
mass and the true mass of the group. The allowed tolerance determines
the `reach' the criterion has in time: larger permitted deviations from virial
equilibrium result in the acceptance of groups that are still collapsing.
We accept groups as real when the virial mass is within 25 per cent of the
true mass.

\subsection{Choice of the galaxy formation parameters}

For the (dimensionless) local density percolation parameters we choose
$p=1$ and $s=\halve$. The maximum percolation length is
$x^{\rm P}_{\rm nn}/2$, which is $0.16 h_0^{-1}$Mpc for our simulations.
We set the upper mass limit for galaxy particles to be
$1.4\times10^{13}$ M$_\odot$. This ensures that possible cD galaxies are
{\it not}\ modelled by single galaxy particles, since that would produce
undesirable numerical problems, and galaxies that massive are not 
(numerically) disrupted anyway. We add an extra limit on the internal
galaxy velocity dispersion of $\sigma_v < 400$ km s$^{-1}$,
the maximum value found for typical ellipticals (de Zeeuw \& Franx 1991)
and comfortably within the velocity dispersion of haloes around spirals
given their typical circular velocities of 200-300 km s$^{-1}$.
Finally, we need to adopt a lower limit of seven particles in a group
because of discreteness noise that causes an artificially
large scatter in the virial mass estimate.

\section{Description and timing of the simulations}

We have run eight simulations of average patches of universe, and
99 cluster models. This latter set forms a catalogue of galaxy clusters,
and is discussed extensively in van Kampen \& Katgert (1997).
The actual N-body code we use is the Barnes \& Hut (1986) treecode,
slightly adapted for our purposes and supplemented with the galaxy
formation algorithm.
We ran the models up to $\sigma_8=1$, which is sufficiently beyond the
time that is expected to be the present epoch for the $\Omega_0=1$ CDM
scenario adopted: $\sigma_8$ was found to be significantly smaller than
unity in most earlier work (e.g.\ Davis et al.\ 1985; Frenk et al.\ 1990;
Bertschinger \& Gelb 1991). 
From a comparison of the galaxy-galaxy autocorrelation function obtained
for the field models to that observed, we find $\sigma_8$ to be in
the range 0.46 to 0.56 (van Kampen 1997), while a similar comparison
of the statistical properties of clusters gives roughly the same range
(van Kampen \& Katgert 1997).

\section{Galaxy properties}

\subsection{Galaxy formation and merging rates}

\begin{figure}
\psfig{file=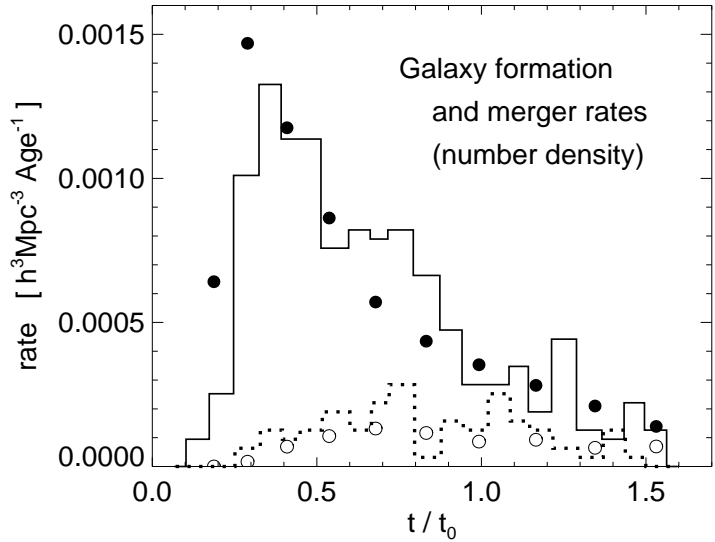,width=10.0cm,silent=}
\caption{Galaxy formation and merger rates for our models (symbols) and
the simulations of Summers (1993, histograms), rescaled to our units and galaxy
masses (see text). Filled symbols and solid lines represent galaxy formation,
open symbols and dotted lines merging.} \label{fig-1}
\end{figure}

As a first check how our modelling of the formation and merging of galaxies
compares to other techniques, notably hydrodynamical simulations, we look at
the galaxy formation and merger number density rates as a function of time.
These are plotted in Figure~\ref{fig-1} for $\sigma_8=0.46$, where $t_0$
is the present epoch. The formation rate peaks at $z\approx 1.3$, whereas
the merger rate does not show a clear peak. The merging of small objects
into galaxies with masses that are included in the formation rate is not
included in the merger rate.

The shapes and amplitudes of both rates compare remarkably well with those
found from hydrodynamical simulations performed by Summers (1993), also
shown in Figure~\ref{fig-1}, if we triple his time-scale. This can be
justified quantitatively as follows: Summers (ibid.) has a higher mass
resolution and forms galaxies down to a lower mass cut-off.
The smallest galaxy masses in his simulations are roughly
a hundred times smaller than our lowest mass galaxies.
The CDM spectrum on galactic scales is a power-law with index -2. We can
then use the scaling law $t_{\rm form}\sim M^{1/4}$ which applies for such a
spectrum to find that this mass difference
gives a factor of three difference in the formation time.

Summers (ibid.) used a full-fledged hydro code (and identified galaxies
with the ordinary friends-of-friends algorithm).
The fact that we find similar shapes and
amplitudes for the rates means that the use
of a galaxy formation recipe with an ordinary collisionless N-body
code can give comparible results to more advanced simulation techniques that
incorporate more (but certainly not all) physical processes.

\subsection{Cluster luminosity function}

\begin{figure}
\psfig{file=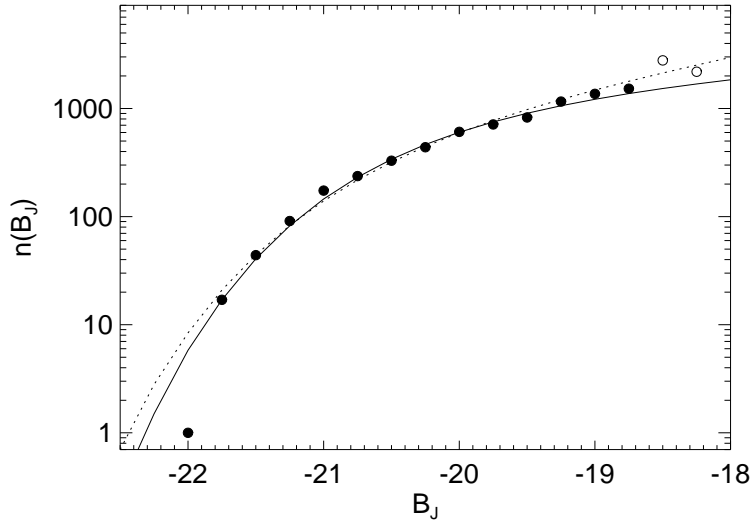,width=10.0cm,silent=}
\caption{Joint luminosity function for all galaxies within the
projected Abell radius. Filled symbols represent galaxies with masses larger
than $1.5\times10^{12}$M$_\odot$. The dashed line indicates a fit of a
Schechter (1976) function to all galaxies. The solid line is a similar fit to
the massive ones only (filled symbols). Both fits were made with $\alpha$ and
$B_{\rm J}^*$ as free parameters, and assume a constant mass-to-light ratio
$\Upsilon_{\rm J}=300$.} \label{fig-2}
\end{figure}

For our cluster simulations, we study the luminosity function within the
projected Abell radius. Since we know only the masses of the galaxies we
need to assume a constant mass-to-light ratio $\Upsilon$ to obtain a luminosity
function. For the $B_{\rm J}$ magnitude, $\Upsilon_{\rm J}\approx1200$ for
an $\Omega_0=1$ universe.
Because on average 25 per cent of the mass is locked into our galaxies
(including dark haloes), $\Upsilon_{\rm J}\approx 300$ for the galaxies.
The joint luminosity function for our cluster models is plotted in
Figure~\ref{fig-2} for all
galaxies (all symbols), and for the $M>1.5\times10^{12}$ M$_\odot$ ones
(filled symbols only), along with a fitted Schechter (1976) luminosity
function for each of them (dashed line for all galaxies, solid line for
the limited set), with slope $\alpha$ and characteristic magnitude
$B_{\rm J}^*$ as free parameters.

The fit for all galaxies is quite good but not perfect: $\alpha=-1.5$.
We find that the fit gets better for the limited set of massive galaxies:
$\alpha=-1.25$. This means that we either do not model low-mass galaxies
very well, or that the mass-to-light ratio is not constant. The first
option is probably true anyway since we cannot model merging
of galaxies with masses below our lower limit towards the low-mass end of
the mass function that we try to fit.

With this in mind it is fair to say that the Schechter function does
fit rather well for the limited set.
We find $B_{\rm J}^*=-20.3$ for this set, which corresponds remarkably
well with the value that Colless (1989) found for a sample of 14 observed
clusters. It compares less well with $B_{\rm R}^*=-22.6$ found by 
Vink \& Katgert (1994) for a sample of 80 clusters, corresponding to
$B_{\rm J}^*\approx-20.8$.
Still we can say that the modelling performes reasonably
well given the uncertainties in both the fits and the assumption of a 
constant mass-to-light ratio.

\acknowledgments

Joshua Barnes and Piet Hut
are gratefully acknowledged for allowing use of their treecode, Edmund
Bertschinger and Rien van de Weygaert for their code to generate initial
conditions, and Eric Deul for allowing me to use the computer systems that are
part of the DENIS project. I acknowledge EelcoSoft Software Services for
partial financial support during the early stages of the project, and an
European Community Research Rellowship as part of the HCM programme during
its final stages.

\end{document}